
\documentstyle{amsppt}

\define\g{\gamma}

\redefine\l{\lambda}

\define\RM{\Bbb R}

\define\gm{\bold g}

\define\<#1,#2>{\langle #1,#2\rangle}
\define\TR{\text{tr}}
\define\dep(#1,#2){\text{det}_{#1}#2}
\magnification\magstep1
\topmatter\title REGULARIZATION OF CURRENT ALGEBRA \endtitle
\author Jouko Mickelsson \endauthor
\affil Department of Mathematics, University of Jyv\"askyl\"a,
SF-40100, Finland\endaffil
\endtopmatter
\NoBlackBoxes
\document
\baselineskip 24pt

ABSTRACT In this talk I want to explain the operator substractions needed
to regularize gauge currents in a second quantized theory. The case of
space-time dimensions $3+1$ is considered in detail. In presence of
chiral fermions the regularization effects a modification of the
local commutation relations of the currents by local Schwinger terms.
In $1+1$ dimensions one gets the usual central extension (Schwinger term
does not depend on background gauge field) whereas in $3+1$ dimensions
one gets an anomaly linear in the background potential.
\footnote""{Presented at the conference "Constraint Theory and Quantization"
in Montepulciano, June 1993}

\vskip 0.4in
1. INTRODUCTION

\vskip 0.4in

Chiral fermions in a nonabelian external gauge field are quantized as
follows. Let $G$ be a compact gauge group, $\gm$ its Lie algebra, $M$ the
physical space, and $\Cal A$ the space of smooth $\gm$ valued vector potentials
in $M.$ For each $A\in\Cal A$ one constructs a fermionic Fock space $\Cal F_A$
containing a Dirac vacuum $\psi_A$. The Hilbert space $\Cal F_A$ carries an
irreducible representation of the canonical anticommutation relations (CAR)
$$a^*(u) a(v) + a(v) a^*(u)= (u,v) \text{ all other anticommutators $=0$}.$$
The representation is characterized by the property
$$a^*(u)\psi_A=0= a(v)\psi_A  \text{ for $u\in H_-(A)$ and $v\in H_+(A)$}
\tag1.1$$
where $H_+(A)$ is the subspace of the one-particle fermionic Hilbert space $H$
spanned by the eigenvectors of the Dirac-Weyl Hamiltonian
$$D_A =i \g_k (\nabla_k + A_k) \tag1.2$$
belonging to nonnegative eigenvalues and $H_-(A)$ is the orthogonal complement
of $H_+(A).$ Here $\nabla_k$'s are covariant derivatives in directions given
by a (local) orthonormal basis, with respect to a fixed Riemannian metric on
$M.$ In the following we shall concentrate to the physically most
interesting case dim$M=3$ and the $\g$-matrices can be chosen as the Pauli
matrices $\sigma_1,\sigma_2,\sigma_3$ with $\sigma_1 \sigma_2= i\sigma_3$
(and similarly for cyclic permutations of the indices) and $\sigma_k^2=1.$

The group $\Cal G=Map(M,G)$ of smooth gauge transformations acts on $\Cal A$
as $g\cdot A= gAg^{-1} + dg g^{-1}.$ The Fock spaces $\Cal F_A$ form a vector
bundle over $\Cal A.$ A natural question is then: How does $\Cal G$ act in
the total space $\Cal F$ of the vector bundle? Since the base base $\Cal A$ is
flat there obviously is a lift of the action on the base to the total space.
However, we have the additional physical requirement that
$$\hat g\hat D_A \hat g^{-1} = \hat D_{g\cdot A} \tag1.3$$
where $\hat D_A$ is the second quantized Hamiltonian and $\hat g$ is the lift
of $g$ to $\Cal F.$  This condition has as a consequence that $\hat g\psi_A$
should be equal, up to a phase, to the vacuum $\psi_{g\cdot A}.$

A complication in all space-time dimensions higher than $1+1$ is that the
representations of CAR in the different fibers of $\Cal F$ are inequivalent,
[A].
The effect of this is that a proper mathematical definition of the
infinitesimal
generators of $\Cal G$ (current algebra) involves further renormalizations
in addition to the normal ordering prescription. In one space dimensions the
situation is simple. The current algebra is contained in a Lie algebra
\define\gl{\bold{gl}} $\gl_1$ which by definition consists of all bounded
operators $X$ in $H$ satisfying $[\epsilon, X]\in L_2,$ where $\epsilon$ is
the sign operator $\frac{D_0}{|D_0|}$ associated to the free Dirac operator
and $L_2$ is the space of Hilbert-Schmidt operators. In general, we denote
by $L_p$ the Schatten ideal of operators $T$ with $|T|^p$ a trace-class
operator. Let $a_n^*=a^*(u_n)$, where $D_0 u_n =\l_n u_n$ and the eigenvales
are indexed such that $\l_n\geq 0$ for $n\geq 0$ and $\l_n<0$ for $n<0.$
Denoting the matrix elements of a one-particle operator $X$ by $(X_{nm})$,
the second quantized operator $\hat X$ is
$$\hat X= \sum X_{nm} :a^*_n a_m : \tag1.4$$
where the normal ordering is defined by
$$ : a^*_n a_m: =\cases a_m a^*_n \text{ if $n=m < 0$ }\\
                 a^*_n a_m  \text{ otherwise }.\endcases$$

The commutation relations are
$$[\hat X,\hat Y]= \widehat{[X,Y]} +c(X,Y) \tag1.5$$
where $c$ is the Lundberg's cocycle, [L],
$$c(X,Y)= \frac14 \TR\epsilon [\epsilon,X][\epsilon,Y].\tag1.6$$
When $X,Y$ are infinitesimal gauge transformations on a circle
the right-hand-side
is equal to the central term of an affine Kac-Moody algebra, [PS],
$$c(X,Y)= \frac{i}{2\pi} \int_{S^1} \TR X' Y.\tag1.7$$
In this talk I want to explain the regularizations needed in $3+1$ space-time
dimensions and the generalization of (1.4) through (1.7).

\vskip 0.4in
2. ACTION OF THE GROUP OF GAUGE TRANSFORMATIONS IN THE FOCK BUNDLE

\vskip 0.4in
Let $\epsilon(A)=\frac{D_A}{|D_A|};$ if $D_A$ has zero modes define
$\epsilon(A)$ to be $+1$ in the zero mode subspace.
For $A\in\Cal A$ denote by $P_A$ the set of unitary operators $h: H\to H$
such that
$$ [\epsilon, h^{-1} \epsilon(A) h]\in L_2.\tag2.1$$
If $h\in P_A$ then also $hs\in P_A$ for any $s\in U_1,$ where $U_1$ is the
group of unitary operators $s$ with the property $[\epsilon,s]\in L_2.$
The spaces $P_A$ form a principal bundle over $\Cal A$ with the structure
group $U_1.$

Since $\Cal A$ is flat the bundle $P$ is trivial and we may choose a section
$A\mapsto h_A\in P_A.$ Define
$$\omega(g;A)= h_{g\cdot A}^{-1} T(g) h_A\tag2.2$$
where $T(g)$ is the one-particle representation of $g\in\Cal G.$ By
construction,
$\omega$ satisfies the 1-cocycle condition
$$\omega(gg';A)= \omega(g;g'\cdot A)\omega(g';A).\tag2.3$$
Using $T(g) D_A T(g)^{-1} = D_{g\cdot A}$ we get $T(g)\epsilon(A) T(g)^{-1}=
\epsilon(g\cdot A)$ which implies
$$\align h_{g\cdot A} [\epsilon,\omega(g;A)]h_A^{-1}& = (h_{g\cdot A}\epsilon
h_{
g\cdot A}^{-1}) T(g) - T(g) (h_A\epsilon h_A^{-1})\\ & \equiv
\epsilon(g\cdot A) T(g)- T(g)\epsilon(A) \text{ mod $L_2$ }
=0.\endalign $$
Since $L_2$ is an operator ideal this equation implies
$$[\epsilon, \omega(g;A)]\in L_2. \tag2.4$$
Thus the 1-cocycle $\omega$ takes values in the group $U_1.$

The group valued cocycle $\omega$ gives rise to a Lie algebra cocycle $\theta$
by
$$\align \theta(X;A)&= \frac{d}{dt} \omega(e^{tX};A)\vert_{t=0}\\
&= h_A^{-1} dT(X) h_A- h_A^{-1}\Cal L_X h_A.\tag2.5\endalign $$
It satisfies the Lie algebra cocycle condition
$$\theta([X,Y];A) -[\theta(X;A),\theta(Y;A)] +\Cal L_X\theta(Y;A)-
\Cal L_Y\theta(X;A) =0,\tag2.6$$
where $\Cal L_X$ is the Lie derivative in the direction of the infinitesimal
gauge transformation $X$, $\Cal L_X f(A)= \frac{d}{dt} f(e^{tX}\cdot A)\vert_
{t=0}.$   We denote by $dT$ the Lie algebra representation in $H$ corresponding
to the representation $T$ of finite gauge transformations. For each $A\in \Cal
A$ and $X\in Map(M,\gm)$ the operator $\theta(X;A)\in \gl_1.$

The section $h_A$ of $P$ can be used to trivialize the bundle of Fock spaces
over $\Cal A.$ Each fiber $\Cal F_A$ is identified as the free Fock space
$\Cal F_0.$ The Hamiltonian $D_A$ is quantized as
$$\hat D_A = q(h_A^{-1} D_A h_A),\tag2.7$$
that is, we first conjugate the one-particle operator $D_A$ by $h_A$ and then
canonically quantize $h_A^{-1}D_A h_A.$  The conjugated operator has a Dirac
vacuum $\psi_A$ contained in $F_0$ (but differing from the free vacuum
$\psi_0$).
The CAR algebra in the background $A$ is represented in $\Cal F_0$ through the
automorphism $a^*(u)\mapsto a^*_A(u)=a^*(h_A^{-1} u),$ $a(u)\mapsto a_A(u)=
a(h_A^{-1}
u)$ and using the free CAR representation for the operators on the right. The
Hamiltonian $\hat D_A$ is then
$$\hat D_A= \sum : a_A^*(u_n) a_A(u_n):\tag2.8$$
where the $u_n$'s for nonnegative (negative) indices are the eigenvectors of
$D_A$ belonging to nonnegative (negative) eigenvalues. The normal ordering is
defined with respect to the free vacuum.

Sections of the Fock bundle are now ordinary $\Cal F_0$ valued functions.
The effect of an infinitesimal gauge transformation consists of two parts:
The Lie derivative $\Cal L_X$ acting on the argument $A$ of the function and
an operator acting in $\Cal F_0,$
$$\hat X = \Cal L_X + \sum  \theta(X;A)_{nm} : a^*_n a_m :, \tag2.9$$
where the $\theta(X;A)_{nm}$'s are matrix elements of $\theta(X;A)$ in the
eigenvector basis $(v_n)$ of $D_0.$
The commutation relations of the second quantized operators are modified by
the Lundberg's cocycle, [M1],
$$[\hat X,\hat Y]= \widehat{[X,Y]} + c(\theta(X;A),\theta(Y;A)).\tag2.10$$
In the next section we want to compute the right-hand side of (2.10) more
explicitly. We shall denote by $c_n(X,Y;A)$ ($n$=dim$M$) the second term on the
right. It is a Lie algebra 2-cocycle in the following sense:
$$c_n([X,Y],Z;A) + \Cal L_X c_n(Y,Z;A) + \text{ cyclic perm. } =0.$$

\newpage
3. A COMPUTATION OF THE COCYCLE

\vskip 0.4in
First we shall construct the section $h_A$ explicitly as a function of the
vector potential when dim$M=3$. We shall define $h_A$ through its symbol, as a
pseudodifferential operator (PSDO) in the spin bundle over $M.$ I claim
that an operator with the following asymptotic expansion satisfies
the requirement (2.1):
$$h_A= 1 + \frac{i}{4}\frac{[\xi, A]}{|\xi|^2} + \text{ terms of lower order
in $|\xi|$}. \tag3.1$$
Here $\xi=\sum \xi_k \sigma_k$ is the three-momentum; its components
represent partial derivatives $-i\partial_k$ in $M,$ with respect to some
local coordinates. In order to make the discussion as simple as possible we
assume that $M$ is the one-point compactification of $\RM^3$ and we use
standard coordinates in $\RM^3.$  We also use the notation $A=\sum A_k
\sigma_k.$

An example of an unitary operator with the asymptotic expansion (3.1) is
the operator
$$h_A=\exp (\frac{i}{4}(D_0^2 +\l)^{-1/2} [D_0,A] (D_0^2 +\l)^{-1/2})
\tag3.2$$
where we have added a small positive constant $\l$ to the denominator
in order to cancel the infrared singularity at $\xi=0;$ this has an
effect in the asymptotic expansion only on terms of order -2 and lower
in the momentum $\xi.$ It is clear that  the lower order
terms do not have any effect on the condition (2.1) since any operator
of order $\leq -2$ is automatically Hilbert-Schmidt when the dimension of
$M$ is 3. Thus we have
$$\theta(X;A)=h_A^{-1}dT(X) h_A -h_A^{-1}\Cal L_X h_A =
X + \frac{i}{4}\frac{[\xi,A]}{|\xi|^2} + O(-2)\tag 3.3$$
where $O(-p)$ denotes terms of order $\leq -p.$ The symbol of the PSDO
$\epsilon$ is $\frac{\xi}{|\xi|}$ and it is a simple computation to check that
indeed $[\epsilon, \theta(X;A)] \in L_2$ using the
product rule of symbols,
$$ (p*q)(\xi,x)= \sum_n \frac{(-i)^{|n|}}{n!} \partial_{\xi}^n p\partial_x^n q
\tag3.4$$
where the sum is over multi-indices $n=(n_1,n_2,n_3)\in \Bbb N^3,$ $|n|=
n_1 +n_2 +n_3$, $n!=n_1!n_2!n_3!$ and $\partial^n_x=
(\frac{\partial}{\partial x_1})^{n_1}(\frac{\partial}{\partial x_2})^{n_2}
(\frac{\partial}{\partial x_3})^{n_3}.$

The term of order -2 in $\theta$ is important in computing the actual value of
$\theta.$ It is equal to
$$\align \theta_{-2} =&-\frac14 \frac{[\sigma_k,A]}{|\xi|^2}\partial_k X
                +\frac12 \frac{[\xi,A]}{|\xi|^4}\xi_k\partial_k X\\
               & +\frac{1}{16}\frac{[\xi,A]}{|\xi|^4}
[\xi,dX].\tag3.5\endalign$$
Note that all terms are linear in the vector potential $A.$
The computation of $c_3(X,Y;A)=c(\theta(X;A),\theta(Y;A))$ is greatly
simplified when we keep in mind that it is only the cohomology class of the
cocycle $c_3$ we are interested in. Another simplification is the following:
Formally,
$$\frac14\TR\epsilon[\epsilon,P][\epsilon,Q]=-\frac12\TR[\epsilon,P]Q\tag3.6$$
when $P,Q$ are in $\gl_1.$ However, the operator on the right is not quite
trace-class; only its diagonal blocks are trace-class. For this reason
the trace is only conditionally convergent. It is convergent when evaluated
with respect to a basis compatible with the polarization $H=H_+\oplus H_-,$
for example, one can choose a basis of eigenvectors of $D_0.$
The trace of an operator $P$ with symbol $p(\xi,x)$ on a $n$-dimensional
manifold is
$$\TR P= (\frac{1}{2\pi})^n\int_{\xi,x} \TR\, p(\xi,x) d^n{\xi} d^n x \tag3.7$$
Note that $P$ is
a trace-class operator iff the order of its principal symbol is less or equal
to -1-dim$M.$

As an exercise, let us compute (3.6) when $M=S^1$ and $P,Q$ are multiplication
operators (infinitesimal gauge transformations). In that case the symbols are
just smooth functions of the coordinate $x$ on the circle. Now $\epsilon=\frac
{\xi}{|\xi|}$ is a step function on the real line, its derivative is twice the
Dirac delta function located at $\xi=0.$  It follows that the symbol of the
commutator $\frac12[\epsilon, P]$ is
$$(-i)\delta_{\xi} p'(x) + \frac{(-i)^2}{2!} \delta'_{\xi} p''(x) +\dots.$$
Applying the formula (3.7) to (3.6) we get
$$\frac14\TR\,\epsilon[\epsilon,P][\epsilon,Q]= \frac{i}{2\pi}\int_{S^1}
\TR\, p'(x)q(x)dx, $$
where the trace under the integral sign is an ordinary matrix trace. If one
feels uneasy with singular symbols, one can approximate $\epsilon$ by a
differentiable function $\frac{\xi}{|\xi|+\l}$ and at the very end let
$\l\mapsto 0.$

In the 3-dimensional case we have to insert $P=\theta(X;A), Q=\theta(Y;A)$ in
(3.6). Using the asymptotic
expansions for $P$ and $Q,$ $p=\sum p_{-k}(\xi,x)$ one has
$$c_3(X,Y;A)= \sum_{j,k} \TR [\frac{\xi}{|\xi|},p_{-j}]*q_{-k}\tag3.8$$
In fact, one needs to take into account only finite number of terms.
The sum of terms with $j+k\geq 4$ is a coboundary of the 1-cochain
$$\sum_{k\geq 4} \TR\left(\epsilon* \theta(X;A)_{-k}\right)\tag3.9$$
Thus we may restrict the sum in (3.8) to indices $j+k <4.$ To take care of the
infrared singularity in the integration in (3.7) we replace all denominators
$|\xi|^{-k}$ by $(|\xi|+\l)^{-k}.$ One can then check by a direct computation
that, modulo coboundaries, the result of the computation in (3.8) does not
depend on the value of $\l$ (i.e., one may take the limit $\l\mapsto 0$ in
cohomology). The final result is in accordance with the cohomological [M, F-S],
[M2], and perturbative arguments, [JJ],
$$c_3(X,Y;A)= \frac{1}{24\pi^2} \int_{M} \TR A [dX,dY].\tag3.10$$

\vskip 0.4in
REFERENCES

\vskip 0.4in
[A] H. Araki in: \it Contemporary Mathematics\rm, vol. 62, American
Mathematical
Society, Providence (1987)

[JJ] R. Jackiw and K. Johnson, Phys. Rev. \bf 182,\rm 1459 (1969)

[L] Lars-Erik Lundberg, Commun. Math. Phys. \bf 50, \rm, 103 (1976)

[M, F-S] J. Mickelsson, Commun. Math. Phys. \bf 97, \rm 361 (1985);
L. Faddeev, S. Shatasvili, Theoret. Math. Phys. \bf 60, \rm 770 (1984)

[M1] J. Mickelsson, Lett. Math. Phys. (1993)

[M2] J. Mickelsson, \it Current Algebras and Groups, \rm Plenum Press,
New York and London (1989)

[PS] A. Pressley and G. Segal, \it Loop Groups\rm, Clarendon Press, Oxford
(1986)

\enddocument